\documentclass{jps-cp}
\usepackage{txfonts} 

\title{Radio Detection of Cosmic Rays -- Achievements and Future Potential}

\author{Tim \textsc{Huege}$^{1}$}

\inst{$^{1}$Karlsruhe Institute of Technology, IKP, P.O. Box 3640, 76021 Karlsruhe, Germany}

\email{tim.huege@kit.edu}

\recdate{April 3, 2017}

\abst{When modern efforts for radio detection of cosmic rays started about a decade ago, hopes were high but the true potential was unknown. Since then, we have achieved a detailed understanding of the radio emission physics and have consequently succeeded in developing sophisticated detection schemes and analysis approaches. In particular, we have demonstrated that the important air-shower parameters arrival direction, particle energy and depth of shower maximum can be reconstructed reliably from radio measurements, with a precision that is comparable with that of other detection techniques. At the same time, limitations inherent to the radio-emission mechanisms have become apparent. In this article, I shortly review the capabilities of radio detection in the very high-frequency band, and discuss the potential for future application in existing and new experiments for cosmic-ray detection.}

\kword{cosmic rays, air showers, radio detection}

\begin{document}
\maketitle

\section{Introduction}

In the quest to solve the mystery of the origin of cosmic rays, it is imperative to collect as much information as possible on each individual detected cosmic-ray event via the combination of complementary detection techniques. In the 1960s it was first discovered that extensive air showers emit pulsed radio signals \cite{JelleyFruinPorter1965}, triggering a decade of intense research \cite{Allan1971}. When progress stagnated, however, the technique was more or less completely abandoned. In the early 2000s, interest was renewed by the availability of powerful digital signal processing techniques, and a number of pioneering projects were initiated. At that time, many promises were made as to the abilities of radio detection of extensive air showers, but the true potential was yet unclear.

In the past decade, tremendous progress has been made in the understanding, measurement and interpretation of radio pulses from cosmic-ray air showers. These developments have recently been reviewed in depth in references \cite{HuegePLREP} and \cite{SchroederReview}, and I do not want to duplicate the content of these review articles here. Instead, I will give a concise overview of the strengths and limitations of radio detection in the very high-frequency band (more specifically, around 30~to~80~MHz), where most of the current experiments operate, and provide a personal view on the potential for future application of radio-detection techniques in the field of cosmic-ray physics.

\section{Capabilities of the radio detection technique}

In this section, I will shortly review the capabilities of radio detection in the very high-frequency range with respect to different aspects. A summary of the discussed points is provided in table \ref{t1}.

\subsection{Understanding of radio emission}

One of the beautiful aspects of the radio detection technique is that the radio signal emitted by extensive air showers can be calculated on the basis of classical electrodynamics. Microscopic Monte Carlo simulation codes devoid of any free parameters such as CoREAS \cite{HuegeARENA2012a} and ZHAireS \cite{AlvarezMunizCarvalhoZas2012} take electron- and positron-tracks simulated by an air shower Monte Carlo simulation as input and from these calculate the radio emission of the complete air shower. Signal coherence is governed by the particle distributions and the light-travel times in the atmospheric refractive index gradient. These microscopic simulations have so far been in agreement with all experimental measurements, both with respect to qualitative features (e.g., the detailed shape of the lateral signal distribution \cite{LOFARNatureXmax}) as well as the absolute signal strength \cite{LOPESrecalibration,TunkaRexInstrument,SLACT510-PRL}. The fact that the latter can be predicted without any possibility of tuning is a strong asset of the detection technique, as it means that the ``radio-emission yield'' is firmly determined from first principles and so radio emission can be used for an absolute calibration of cosmic-ray detectors.

In addition to the microscopic simulation of air showers, macroscopic models (e.g., \cite{ScholtenWernerRusydi2008}) have been developed. They have been imperative to gaining an understanding of the dominant emission mechanisms, namely geomagnetically induced time-varying transverse currents and a time-varying negative charge excess (also called Askaryan effect), which superpose with characteristic signal polarizations leading to an asymmetric signal distribution in the shower plane. 

\subsection{Sensitivity to electromagnetic cascade}

The radio emission purely arises from electrons and positrons in the air shower because for these particles the charge-to-mass ratio is by far the largest. This means that radio measurements directly access the electromagnetic cascade of extensive air showers. This reduces the sensitivity to hadronic interaction models and allows direct comparison of radio measurements to measurements with fluorescence detectors. (There is a very small difference between measurements by radio and fluorescence detectors because a small additional amount of fluorescence light arises from the muonic cascade, which does not contribute to the radio signal. This difference is typically on the 1--2\%-level, though.) Also, combination of radio detectors with muon detectors is a promising approach, especially to address remaining issues in the modelling of the muonic component of air showers.

\subsection{Calorimetric energy measurement}

The radio signal from an extensive air shower is calorimetric in the sense that the energy contained in the coherent radio signal -- the ``radiation energy'' -- scales with the square of the energy contained in the electromagnetic cascade of an air shower \cite{AERAEnergyPRL}. Furthermore, there is no relevant absorption or scattering of radio waves at MHz frequencies in the atmosphere. A radio detector that can sample the lateral signal distribution well enough to perform an area integration and calculate the total energy deposited on the ground can thus determine the energy in the electromagnetic cascade both accurately and precisely. The accuracy is determined mostly by the absolute calibration of the radio antennas, and values as good as 10\% have recently been achieved \cite{Aab:2017lby}. The precision is hardly affected by shower-to-shower fluctuations with an intrinsic resolution of better than 5\% \cite{GlaserRadEnergyStudy}. Actual experimental measurements so far have reached energy resolutions of approximately 15\% \cite{AERAEnergyPRD}, but with potential to go to 10\%.

Another possibility to do energy measurements with radio detectors is to estimate the energy from a radio-signal amplitude measurement at a characteristic lateral distance from the shower axis. This also achieves precisions of approximately 15\% \cite{ApelArteagaBaehren2014,TunkaRexCrossCalibration}, but unlike the radiation energy is not directly comparable between experiments performed at different observation altitudes.

\subsection{Observation duty cycle}

One of the promises of radio detection was that, unlike fluorescence detection, it can be applied 24 hours per day. Indeed, this is true, with the only so-far observed limitation being thunderstorms in the vicinity of the radio-detection array (which in fact can be studied with radio measurements of cosmic-ray induced air showers \cite{LOFARThunderstorms2015}). Typical duty cycles thus reside well above 95\%.

\subsection{Angular resolution}

The angular resolution of radio detection is very high. It is difficult to determine from measurements as the reference direction supplied by other detectors is typically less precise. However, LOFAR measurements \cite{LOFARWavefront2014} as well as simulation studies in the context of LOPES \cite{LinkThesis2016} indicate that an angular resolution of 0.1$^{\circ}$ is achievable. For cosmic-ray detection such high angular resolutions are not needed, but for the detection of photon-induced air showers this could be a strong asset.

\subsection{Mass sensitivity}

There are several imprints of the particle mass on the radio signals from extensive air showers. The one most widely exploited so far is the effect that the relativistically forward-beamed radio emission subtends larger areas with a less steep lateral signal distribution if the shower maximum is geometrically further away \cite{HuegeUlrichEngel2008}, i.e., for iron-induced showers as compared with proton-induced showers. The achievable resolution depends on the density with which the radio-emission footprint is sampled. The dense LOFAR array has achieved an $X_{\mathrm{max}}$ resolution of 17~g/cm$^{2}$ \cite{LOFARNatureXmax}. Sparser arrays such as Tunka-Rex \cite{TunkaRexCrossCalibration} and AERA \cite{GateARENA2016} have so far achieved resolutions of $\sim40$~g/cm$^{2}$ in the energy range from 10$^{17}$ to $10^{18}$~eV. There is still room for further improvement in the reconstruction methods, as so far only the signal strengths are used as input, but not the signal polarization or signal timing, both of which are also known to have sensitivity to $X_{\mathrm{max}}$. Extension of the frequency range to higher frequencies also has the potential to significantly improve the $X_{\mathrm{max}}$ resolution, as has been shown in recent simulation studies in the context of the SKA \cite{AnneSKAARENA2016}.

\subsection{Economics of radio detection}

One of the promises of radio detection has also been to lower the cost of cosmic-ray detection arrays. It is true that the actual detector, the antenna and associated amplifiers, can be built very economically. The SALLA antennas used in Tunka-Rex, for example, can be manufactured at a cost of roughly 500~USD including low-noise amplifier \cite{SchroederUHECR2014}. Digital electronics are needed to sample the signal, but these profit directly from Moore's law and are thus becoming more cost-effective as time passes. A two-channel (two-polarization) radio detector is certainly possible for a price of 1000~USD. There are, however, two limitations. The first is related to the needed detector spacing (see next subsection), the second is related to the ``infrastructure'' needed to operate the radio detectors. This includes in particular the power supply and the data transmission. For data transmission, nowadays powerful and economic options are available off-the-shelf, and with the upcoming LTE-advanced standard this will become even better. Power supply, however, remains a problem. Both the distribution of power with long cables as well as the decentralized generation using battery-buffered photovoltaic systems involves significant costs and/or maintenance needs. An attractive approach is thus to not operate radio detectors in stand-alone mode but combine them with complementary detectors such as particle detectors, so that as much of the infrastructure as possible can be shared.

\subsection{Detector spacing}

One of the limitations of the radio detection technique that became apparent over the past decade is the relatively dense detector spacing needed. This is due to the forward-beamed nature of the radio emission. The area illuminated on the ground is directly related to the geometrical distance of the shower maximum. For near-vertical air showers the shower maximum is close-by and the radio-emission footprint has a diameter of only 100--200~m, within which very high signal amplitudes are present. At a zenith angle of 50$^{\circ}$, the footprint has a diameter of up to 1000~m (in the shower plane, after correcting for projection effects), with on average lower signal strengths. For inclined air showers with zenith angles of 75$^{\circ}$ or more, the radio-emission footprint reaches diameters of 2000~m or more (in the shower plane) \cite{KambeitzARENA2016}, with yet lower average signal strengths \cite{HuegeUHECR2014}. It is important to note that an increase in the cosmic-ray energy will not alter this picture significantly, as the signal distribution falls off very quickly laterally and thus an increase in overall signal strength will only negligibly increase the illuminated area.

In effect, this means that different antenna densities probe different zenith angle ranges and cosmic-ray energies. Dense antenna arrays (spacing below 100~m) are most-suited for near-vertical air showers with relatively low cosmic-ray energies (around and slightly below $10^{17}$~eV). Intermediate spacings of order 200-300~m are most suited for zenith angles of 30$^{\circ}$ to 60$^{\circ}$ and energy ranges of $10^{17}$ to a few times $10^{18}$~eV. Large spacings of order 1000~m or larger are suitable for detection of inclined air showers at energies well beyond $10^{18}$~eV. A graded array design could access all of these energy scales, but each with less than 2$\pi$ solid angle.

\subsection{Detection threshold}

Successful detection of radio emission from extensive air showers is possible once the signal strength is high enough that it is detectable in the presence of radio background (see next subsection). As discussed in the previous subsection there is a direct connection between zenith angle and average signal strength, so that the energy threshold for radio detection also depends on zenith angle. Furthermore, the signal strength scales as $\sin(\alpha)$, where $\alpha$ denotes the so-called geomagnetic angle, the angle between the shower axis and the geomagnetic field axis. In other words, also an azimuth dependence of the signal strength and thus detection threshold arises. The most conservative approach of requiring 100\% detection efficiency from all directions would severely limit statistics of radio measurements. As the directional dependence of the radio signal is, however, well understood, such a radical approach is usually not needed. Instead, the direction-dependent thresholds and biases can be modelled precisely and taken into account in the analysis. Nevertheless, this is of course a drawback and in particular anisotropy studies using radio detection are likely to be challenging.

\subsection{Radio backgrounds}

One difficulty in radio detection is the omnipresence of background noise. This has to be differentiated between continuous and pulsed background noise. In the frequency band from 30~to~80~MHz the baseline continuous background is given by radio emission from the Galaxy \cite{CCIR670}. Successful pulse detection in individual radio antennas typically requires signal strengths of order 1--2 $\mu$V/m/MHz, leading to the aforementioned typical detection thresholds as a function of zenith angle. It might be an option to extend the detection band to higher frequencies, where Galactic noise fades quickly --- however, the radio-emission footprint will then start to show a more prominent Cherenkov ring \cite{Nelles201511}, i.e., the radio array will at the same time need to become denser. In populated areas, also the FM band between 85~and~110~MHz presents a problem.

Pulsed noise from anthropogenic sources (badly insulated power lines, machinery, cars, ...) pre\-sents a challenge for triggering cosmic-ray pulses in the radio-signal data stream itself. Also, reconstruction can be significantly more difficult in the presence of a high number of noise pulses. The challenges posed by anthropogenic noise are the main reason why so far self-triggered radio detection has had limited success. In an environment devoid of anthropogenic noise pulses, however, self-triggering should be feasible, as also demonstrated by ANITA \cite{HooverNamGorham2010} and ARIANNA \cite{Barwick:2016mxm}.

\section{Application potential}

Here, I will shortly discuss the potential I personally see in the application of radio-detection techniques in the context of cosmic-ray physics.

\subsection{Improvement of hybrid measurements}

It is clear that radio measurements provide very valuable information that any existing cosmic-ray detector can profit from. This includes precise, and accurate, measurements of the energy in the electromagnetic cascade of air showers as well as sensitivity to the depth of shower maximum. Generally speaking, these measurements will have different systematic uncertainties than those of other detection techniques. Combination in a hybrid approach will thus significantly improve the measurement quality of each individual air-shower event.

Such a hybrid detection scheme can best be achieved if a significant fraction of the infrastructure for data communication and power distribution can be re-used. Then, the cost for deploying an additional radio detector array is very moderate. A caveat exists in the required antenna spacing. For non-inclined air showers, a spacing of less than 300~m is recommended, so that the accessible energy range is typically around $10^{17}$~eV up to several 10$^{18}$~eV.

\subsection{(Cross-)calibration of the energy scale}

One of the hardest challenges in cosmic-ray physics is the accurate calibration of the absolute energy scale of cosmic rays. Calibrations resting on Monte Carlo simulations of extensive air showers directly inherit the significant systematic uncertainties of hadronic interaction models. The so-far best absolute calibration technique is based on fluorescence detection \cite{EnergyScaleICRC2013}, which however requires enormous efforts in the monitoring and modelling of the atmosphere.

Radio detection could provide very valuable additional information to calibrate the absolute energy scale of cosmic-ray detectors: as discussed earlier, the signal can be predicted from first principles, and the measurement is not influenced significantly by atmospheric conditions. Especially in the energy range from $10^{17}$~eV up to several times $10^{18}$~eV a small radio-detection array can be an accurate and cost-effective way to calibrate or validate the energy scale of a cosmic-ray detector. This has already been demonstrated successfully for a cross-calibration between the KASCADE-Grande and Tunka experiments via their respective radio detector extensions LOPES and Tunka-Rex \cite{Apel:2016gws}.

\subsection{Measurements of inclined air showers}

A radio detector array focusing on inclined air showers would be able to observe the highest energy cosmic rays with an antenna spacing consistent with the typical grid layout of particle detector arrays, i.e., well above 1000~m. Re-use of existing infrastructure would make the deployment of a radio antenna at the location of existing particle detectors very cost-effective. The combined measurement with particle detectors and radio antennas would then allow a good separation between muons (basically the only part of the air shower reaching the ground at high zenith angles) and the electromagnetic cascade (as measured by the radio antennas). In effect, the measurement strategy being followed with the AugerPrime upgrade \cite{AugerPrimeIcrc2015} could be applied to larger zenith angles, providing additional statistics, allowing independent cross-checks, and accessing additional regions of the sky.

\subsection{Precision measurements with dense arrays}

The potential of precision measurements of individual cosmic-ray events has been demonstrated with the dense LOFAR array. Unfortunately, the inhomogeneous distribution of antennas within LOFAR as well as the low effective duty cycles (arising mostly from organizational issues) limit the potential. A future dense radio detection array equipped with cosmic-ray measurement functionality could carry out high-precision measurements in the energy range between $10^{16.5}$ and $10^{18}$~eV. One particular example is the upcoming Square Kilometre Array (SKA) for which an initial simulation study predicts achievable $X_{\mathrm{max}}$ resolutions of better than 10~g/cm$^{2}$ \cite{AnneSKAARENA2016}. This is very encouraging, especially since this study was only applying reconstruction approaches available today, which so far only exploit the signal strength and completely neglect polarization and phase information. In principle, it should even be possible to use radio measurements to do a ``tomography'' of the electromagnetic cascade of the air shower. Detailed studies of air shower physics might be possible with such approaches, which, however, have yet to be developed.

\section{Conclusions}

Radio detection of cosmic rays has matured and progressed tremendously in the past decade. It has many strengths such as the possibility to calculate the signal from first principles, an accurate determination of the energy in the electromagnetic cascade of air showers, and good sensitivity to the depth of shower maximum. Limitations exist in particular in the dimensions of the radio-emission footprint, requiring relatively dense antenna arrays, except for horizontal air showers. Radio arrays thus have to be tailored to a specific energy range, in which they can then yield very useful additional information, in particular in hybrid approaches with particle detectors. Both sparse and dense arrays have significant potential, if chosen appropriately for a given scientific goal.

\begin{table}[tbh]
\caption{This table gives a concise summary of the capabilities of the radio detection technique.}
\label{t1}
\begin{tabular}{p{30mm}p{115mm}}
\hline
understanding of \newline radio emission & signal well-understood, can be predicted from first principles, firm prediction of absolute strength (``yield'') can be used to set energy scale \\
\\[-0.7em]
sensitivity to electromagnetic cascade & radio signals arise purely from electrons and positrons, direct comparison to fluorescence detection possible, small influence of hadronic interactions\\
\\[-0.7em]
calorimetric energy \newline measurement & no absorption or scattering of radio signals in the atmosphere, radiation energy directly related to energy in electromagnetic cascade, $\sigma_{\mathrm{E}}<15$\% demonstrated, $\sigma_{\mathrm{E}}<10$\% seem feasible, absolute (cross-)calibration between detectors using radio seems much simpler than with fluorescence detectors\\
\\[-0.7em]
duty cycle & in principle 100\%, thunderstorms only exception, typical values $>95$\%\\
\\[-0.7em]
angular resolution & $\sigma<0.5^{\circ}$ easily achievable, probably even $\sigma<0.1^{\circ}$\\
\\[-0.7em]
mass sensitivity & so far demonstrated $\sigma_{\mathrm{Xmax}}<20$~g/cm$^{2}$ for dense arrays, $\sigma_{\mathrm{Xmax}}\sim40$~g/cm$^{2}$ for sparse arrays, further potential when using signal polarization and timing, very dense arrays can probably achieve $\sigma_{\mathrm{Xmax}}\sim10$~g/cm$^{2}$\\
\\[-0.7em]
economics of radio \newline detection & USD 1000 per two-channel radio-detector achieved, significant cost in infrastructure for desired antenna spacing, in particular power distribution, hybrid approaches with shared infrastructure are attractive\\
\\[-0.7em]
detector spacing & $<300$~m for $\theta<60^{\circ}$, $>1000$~m for $\theta>70^{\circ}$, low zenith angles probe low energies, high zenith angles probe high energies\\
\\[-0.7em]
detection threshold & zenith-dependent (relativistic beaming) and azimuth-dependent (geomagnetic effect) detection threshold, can be modeled precisely, 100\% detection efficiency difficult to achieve and would severely limit statistics \\
\\[-0.7em]
radio backgrounds & continuous Galactic noise sets detection threshold at $\sim10^{17}$~eV, possibly better signal-to-noise ratio when including higher frequencies or interferometry, pulsed anthropogenic noise affects self-trigger and reconstruction, external trigger more effective than self-trigger when not in very radio-quiet environment\\
\hline
\end{tabular}
\end{table}



\end{document}